\documentclass[prb,aps,twocolumn]{revtex4}

\begin{document}

\title{Comment on \textquotedblleft Superinsulator and Quantum
Synchronization\textquotedblright }
\author{K. B. Efetov$^{1,2}$, M. V. Feigel'man$^2$ and P. B. Wiegmann$^{3,2}$%
}
\affiliation{$^1$Theoretische Physik III, Ruhr-Universitat Bochum, D-44801 Bochum, Germany}
\affiliation{$^2$L. D. Landau Institute for Theoretical Physics, Moscow 119334, Russia}
\affiliation{$^3$James Frank Institute, University of Chicago, IL 60637, USA}
\date{\today}

\begin{abstract}
We show that the \textquotedblleft theory\textquotedblright\ of
\textquotedblleft superinsulating\textquotedblright\ state presented by
V.Vinokur et al (Nature vol. 452, p. 613, 2008) and Fistul et al (Phys. Rev.
Lett. vol. 100, 086805, 2008 ) is essentially incorrect due to a sequence of
errors in the theoretical analysis of the standard model of Josephson arrays
which properties have been carefully studied and described in the
literature. The line of calculations suggested in these articles lead to
unphysical results. In particular, the calculations predict a direct current
flowing through a capacitor. Moreover, this current may flow even in the
absence of voltage - a sort of \textit{supercurrent} flowing in the
"superinsulating" state. We also question that the theoretical model
employed in these works, even if treated correctly, is applicable to the
analysis of experimental data on homogeneously disordered superconductive
films.
\end{abstract}

\maketitle

\textbf{1.} Two recent articles \cite{VBnature,VBprl} (hereafter referred to
as I and II) are supposed to contain theory of \textquotedblleft
superinsulating\textquotedblright\ state with zero linear resistance first
seen in InO$_{x}$ films near superconductor-insulator transition by
Sambandamurthy \textit{et al} \cite{Murthy} and later observed by Baturina
\textit{et al} \cite{BaturinaExp} in thin films of TiN (unfortunately we
found no reference in I to the first observation of this phenomenon ~\cite%
{Murthy}). Calculations presented in I and II are based on the same model
and are almost identical to each other both in the method and results (few
details where they differ will be mentioned below), which allows us to
analyze them in parallel.

Although the materials studied experimentally in Refs. \cite%
{Murthy,BaturinaExp} are disordered homogeneous films, the authors of I and
II suggested to model them by a regular 2D Josephson junctions (JJ) array.
Moreover, they have arbitrarily chosen a special case when the
self-capacitance of the islands $C_{0}$ can be neglected with respect to the
junction capacitance $C$. This case may be relevant to some special
geometries of artificial networks but its relevance to the materials
investigated in Ref. \cite{Murthy,BaturinaExp} has not been justified in the
papers and\textbf{\ }is questionable (as will be discussed at the end of the
present Comment).

Studying the model of the regular JJ array the authors of I, II claim that
they discovered a \textquotedblleft superinsulating" state with a huge
dielectric gap (that grows with the size of the system) and argue that this
state is due to a new phenomenon they called \textquotedblleft quantum
synchronization".

In this Comment, we are not going to present our own interpretation of the
interesting experimental results~of Refs. \cite{Murthy,BaturinaExp}. Our aim
here is to demonstrate that the theoretical part of the paper I, as well as
the paper II, contains serious errors leading to wrong conclusions.
Theoretical content of these papers consists both of calculations (p.2 of II
and \textquotedblleft Methods\textquotedblright\ part of I) and of a number
of qualitative arguments.

We begin with analysis of these calculations and of the main result, Eqs.
(1) in both papers, then we comment on some of these qualitative arguments.
After that we present well-known (for about two decades) results on how the
model studied in I, II really behaves at low temperatures, and finally
discuss an applicability of this particular JJ array model to the
experimental data of Refs.~\cite{VBnature,BaturinaExp}.

\textbf{2.} The main results of I, II are displayed in the form:
\begin{eqnarray}
I &\propto &\exp \left[ -\frac{\left( \Delta _{c}-eV\right) ^{2}\exp
(E_{c}/2T)}{\Delta _{c}E_{c}}\right]  \label{e1a} \\
I &\propto &\exp \left[ -\frac{\left( \Delta _{c}-eV\right) ^{2}}{2\Delta
_{c}T}\right]  \label{e1}
\end{eqnarray}%
and it is stated by the authors of I that Eq. (\ref{e1a}) goes over to Eq. (%
\ref{e1}) at $T\gg E_{c}$ (in fact it does not). Here $I$ is the current
through the system, $V$ is the voltage and $\Delta _{c}$ is the
\textquotedblleft new" gap estimated by the authors as
\begin{equation}
\Delta _{c}=\left\{
\begin{array}{c}
E_{c}\min \left\{ \lambda _{c},L\right\} /d,\text{ for 1D arrays} \\
\left( E_{c}/2\right) \ln \left( \min \left\{ \lambda _{c},L\right\}
/d\right) ,\text{ for 2D arrays}%
\end{array}%
\right.  \label{e2}
\end{equation}%
and $\lambda _{c}=d\sqrt{E_{c0}/E_{c}}\gg d$ is the screening length ($%
E_{c}=e^{2}/2C$, $E_{c0}=e^{2}/2C_{0}$), $L$ is the sample size; $C_{0}$ is
the self-capacitance of each island and $C\gg C_{0}$ is the capacitance of
each junction.

The \textquotedblleft superinsulating\textquotedblright\ behavior claimed to
be found in I consists of a very fast (double-exponential) temperature
decrease of the conductance at low voltage $eV<\Delta _{c}$ and low
temperatures $T<E_{c}$, cf. (\ref{e1a}). Eq. (\ref{e2}) describes a huge
(especially for 1D case) gap $\Delta _{c}$ depending on the size of the
system and the authors argue that it is formed due to the phenomenon of the
\textquotedblleft quantum synchronization". In fact, as we show below, Eqs. (%
\ref{e1a}, \ref{e1}) cannot be derived from the suggested model.

The Lagrangian $L$ of the problem is defined by Eqs. (3,4) in I and by Eq.
(4) in II. With a small modification, keeping only relevant terms considered
in I and II, this Lagrangian can be written as
\begin{equation}
L=L_{0}+L_{L}+L_{R}+L_{LR}  \label{e7}
\end{equation}%
where
\begin{eqnarray}
L_{0}&=&\sum_{ij,kl}\left[ \frac{\hbar ^{2}}{4E_{c}}\left( \dot{\chi}_{ij}-%
\dot{\chi}_{kl}\right) ^{2}+E_{J}\cos \left( \chi _{ij}-\chi _{kl}\right) %
\right]  \nonumber \\
&&+\sum_{ij}\frac{\hbar ^{2}}{4E_{c0}}\dot{\chi}_{ij}^{2}  \label{e8}
\end{eqnarray}%
and
\begin{equation}
L_{L}=E_{J}\sum_{j=1}^{M}\cos \left( eVt/\hbar +\chi _{1j}\left( t\right)
+\psi \left( t\right) \right)  \label{e9}
\end{equation}%
\begin{equation}
L_{R}=E_{J}\sum_{j=1}^{M}\cos \left( -eVt/\hbar +\chi _{Nj}\left( t\right)
-\psi \left( t\right) \right)  \label{e10}
\end{equation}%
\begin{equation}
L_{RL}=\frac{\hbar ^{2}}{8E_{c}}\sum_{j=1}^{M}\left[ \dot{\chi}_{1j}+\dot{%
\chi}_{Nj}\right] ^{2}  \label{e11}
\end{equation}%
The phases $\chi _{1j}$ and $\chi _{Nj}$ are the leftmost and rightmost
phases of the sample (next to the leads).

The authors of I and II write explicitly that, when deriving Eq. (\ref{e1}),
they neglect all Josephson couplings inside the array, i.e. terms with $%
E_{J} $ in $L_{0}$. (cf. paragraphs between Eqs. (12) and (13) in I and
between Eqs.(8) and (9) in II). 

However, once this 
approximation 
is made, 
the system becomes electrically disconnected for \textit{dc} current between
its left and right terminals. Neglecting the Josephson coupling $E_{J}$ we
see that the right and left parts of the system are connected to each other
by the Coulomb interaction only (first term in Eq. (\ref{e8})), which means
that the only allowed current is the capacitive one, $I(t)\propto C_{\mathrm{%
eff}}dV(t)/dt$, which is strictly zero in the \textit{dc} limit. On a more
formal level, the neglect of all the intrinsic Josephson couplings leads to
the possibility of rotating the phase difference $\chi _{1j}(t)-\chi
_{Nj}(t) $ by an arbitrary constant without any energy cost (gauge
invariance). Writing the corresponding expression for the current and
integrating over static $\chi _{1N}$ or $\chi _{Nj}$ in the interval $%
(0,2\pi )$ one comes to the strictly zero \textit{dc} current.

How could it happen that the authors of I, II got the nonzero direct
current? We believe that it is due to an incorrect approximation made during
the derivation of Eqs. (11) and (7) of the articles I and II, respectively.
Namely, they first represent the sum of two cosine terms $L_{R}+L_{L}$ in
the equivalent form of the product (and then the same with the expression
for the current):
\begin{eqnarray}
L_{R}+L_{L} &=&2E_{J}\sum_{j=1}^{M}\cos \left[ \frac{\chi _{1,j}+\chi _{N,j}%
}{2}\right] \\
&&\times \cos \left[ \frac{2eVt+\psi +\chi _{1,j}-\chi _{N,j}}{2}\right]
\nonumber  \label{product}
\end{eqnarray}%
and then they replace the\textit{\ average of the product} of two nontrivial
functions of time (e.g. Eq.(7) of I) by the \textit{product of their averages%
}, thus coming to Eq. (11) of I (or Eq.(7) of II). We rewrite these
equations for the current $I_{s}\left( V\right) $ here for an easy reference
\begin{equation}
I_{s}\left( V\right) =A\mbox{Im}\int_{0}^{\infty }dt\exp \left[ -\delta
t+2ieVt\right] K\left( t\right)  \label{a1}
\end{equation}%
where $A$ is a constant proportional to $E_{J}^{4}$, and the function $%
K\left( t\right) $ equals%
\begin{equation}
K\left( t\right) =\left\langle \exp i\left[ \chi _{1}\left( t\right) -\chi
_{1}\left( 0\right) -\chi _{N}\left( t\right) +\chi _{N}\left( 0\right) %
\right] \right\rangle _{L_{0}}  \label{a2}
\end{equation}%
and $t$ is real time.

This procedure of I and II could make sense under some special
circumstances. Indeed, in the situation studied in Refs. [18,19] of I, there
are no intrinsic array junctions and the autocorrelation function $K(t)=1$.
The same approximation could also be used if the Josephson couplings of the
bulk were different from those near the leads and were sufficiently large
(i.e. if the condition $E_J \gg E_c$ would be fulfilled in the bulk of the
array). In this case, phase fluctuations in the bulk would be strongly
suppressed and this situation would not be essentially different from the
one considered in Refs. [18,19]. This means that Eq. (\ref{a1}) can in some
cases (when $K\left( t\right) $ is slow function of time) be correct and the
calculation based on the analog of the representation (I.11) is, in
principle, possible.

However, an attempt of Vinokur \textit{et al} (I) and Fistul \textit{et al}
(II) of using formulae written in [18,19] to their own model failed
explicitly due to the neglect of cross-correlations between two
trigonometric factors contained in Eq.7 of I \textit{for vanishing values of
the Josephson couplings in the bulk}. The function $K\left( t\right) $
calculated in I, II changes with time extremely fast (on the scale of $%
\Delta _{c}^{-1}$) and there is no justification of the "decoupling"
approximation employed.
The correct calculation \textit{within their approximation of zero Josephson
couplings inside the sample} would lead the authors to the evident and
trivial result: \textit{dc} current cannot flow through a capacitor.

Furthermore, Eqs. (\ref{e1a}, \ref{e1}) were written in I, II without
specifying the form of the pre-exponential. Following the calculation of I,
II one comes to the conclusion that the pre-exponential is proportional to $%
E_{J}^{4}$ (see Eqs. (\ref{a1}, \ref{a2})). This is incorrect: expanding in
the coupling terms of $L_{0}$ one can see that the coefficient in front of
such a term is strictly zero if $N>4$. In fact, using the initial
Lagrangian, Eqs. (\ref{e7}-\ref{e11}), a non-zero contribution to the
current could be obtained (neglecting all inelastic processes as the authors
of I and II did) in the $N$-th order of the expansion in a small parameter $%
E_{J}/E_{c}$ only (where $N$ is the size of the array).

One could wonder, however, if Eqs. (1,2) could still be derived, although
with a different coefficient (not just proportional to $E_{J}^{4}$).

We found that Eqs. (\ref{e1a}, \ref{e1}) cannot be repaired in such a way.
The form of Eqs. (\ref{e1a}) and (\ref{e1}) leads to the exponential
suppression of the current not only at low voltage $eV<\Delta _{c}$ but at $%
eV>\Delta _{c}$ as well. This surprising phenomenon was not commented in
paper II. At the same time, in the later paper I the authors mentioned that
they used their result (1) for $eV<\Delta _{c}$ only. However, such a
limitation does not follow from any step made in their derivation of this
result and the need to invoke it on a purely verbal level indicates
erroneous nature of the analysis.

Another problem occurs when the authors neglect contributions of non-zero
winding numbers at $T$ larger than $E_{c}$ in Eq. (I.13). The quantization
of charge is inherently related to the summation over all winding numbers.
No insulating behavior can be obtained without it.

Actually, the necessity of taking into account the discreteness of the
charges even for $T>E_{c}$ is specific for the model with the vanishing
self-capacitance $C_{0}$ considered in the articles I, II. Adding the
elementary charge $2e$ to the system costs the huge energy $\Delta _{c}$,
Eq. (\ref{e2}), and not $E_{c}$. Therefore, one \textit{must} sum over the
integer charges for all temperatures $T<\Delta _{c}$ and the substitution of
the sum by an integral is impossible. In terms of actual calculations it
means that the integral over non-zero Fourier-components $\chi _{\omega }$
\textit{must} be complemented by the sum over nonzero integer winding
numbers.

Neglecting winding numbers the authors take the Gaussian integral over the
phase $\chi $ and obtain (see Eq. (10) of II).
\begin{equation}
K\left( t\right) =\exp \left( -2\Delta _{c}E_{c}\zeta \left( T\right)
t^{2}-2i\Delta _{c}t\right)  \label{e12}
\end{equation}%
where $\zeta \left( T\right) =T/E_{c}$ for $T>E_{c}$.

Note the presence of $t^{2}$ term in Eq. (\ref{e12}). It is this term that
might lead to Gaussian integrals and, eventually, to Eqs. (\ref{e1a}, \ref%
{e1}) for some physical quantities. However, the summation over the non-zero
winding numbers $k$ results in the vanishing of the $t^{2}$ term in the
exponent in Eq. (\ref{e12}). This is explicitly demonstrated, e.g., in Ref.
\cite{ET}, where a granular metal consisting of normal grains was
considered. In the absence of the $t^{2}$ term in the exponent in the
function $K\left( t\right) $ one would not be able to get generally the
combination $\left( \Delta _{c}-eV\right) ^{2}$ in the exponents in Eqs. (%
\ref{e1a}, \ref{e1}).

There is an exception when the correlation function $K\left( t\right) $ can
really be written in a Gaussian form for\textbf{\ }$T<\Delta _{c}$. This is
the case of a one dimensional chain of JJ with the vanishing
self-capacitance $C_{0}$. In this case, the correlation function $K\left(
t\right) $, Eq. (\ref{a2}), factorizes into a product of correlation
functions for each junction between the grains. The authors of I, II use
this property and bring the expression for the correlation function $K\left(
t\right) $ to the form of Eq. (\ref{e12}) for arbitrary temperature with a
function $\zeta \left( T\right) $ proportional to $T$ at $T>E_{c}$, and to $%
\exp \left( -E_{c}/T\right) $ at $T<E_{c}$. Then, apparently calculating
Gaussian integrals they come to Eqs. (\ref{e1a}, \ref{e1}), thus obtaining
the double-exponential behavior at low temperatures. The latter is supposed
to describe the \textquotedblleft superinsulator".

However, direct inspection shows that Eqs.(\ref{a1}, \ref{e12}) do not lead
to the results announced in Eqs.(\ref{e1a}, \ref{e1}) (at this moment we do
not even discuss the validity of (\ref{a1}, \ref{e12})).

Substituting Eqs. (\ref{e12}) into Eq. (\ref{a1}) (Eqs. (14) or (15) of I
into Eq. (11) of I) we obtain
\begin{equation}
I_{s}\left( V\right) \!=\!A\int_{0}^{\infty }\!\sin \left[ 2\left(
V\!-\!\Delta _{c}\right) t\right] \exp \left( -\delta t\right) K_{0}\left(
t\right) dt,  \label{a7}
\end{equation}%
where $K_{0}(t)$ is a real, even function of time defined via the relation $%
K\left( t\right) =\exp \left( -iNE_{c}t/2\right) K_{0}\left( t\right) $.

We see immediately from Eq. (\ref{a7}), that
\begin{equation}
I_{s}\left( V\right) \neq -I_{s}\left( -V\right)  \label{a8}
\end{equation}%
In particular, Eq.(\ref{a8}) shows that the \textquotedblleft
superinsulating" state can carry a non-zero current without any voltage or
magnetic field applied, i.e. the ground state maintains a \textquotedblleft
supercurrent":
\begin{equation}
I_{s}\left( 0\right) =-A\int_{0}^{\infty }\sin \left[ \left( 2\Delta
_{c}t\right) \right] \exp \left( -\delta t\right) K_{0}\left( t\right) dt
\label{a9}
\end{equation}%
It is not difficult to see that this supercurrent is even not\textbf{\ }%
small. Using Eq. (\ref{e12}) we obtain in the limit $\delta \rightarrow 0$%
\begin{equation}
I_{s}\left( 0\right) \simeq -A/\Delta _{c}  \label{a10}
\end{equation}%
The only way to obtain Eqs. (\ref{e1a}, \ref{e1}) from Eqs. (\ref{a1}, \ref%
{e12}) is to use $\mbox{Im}K\left( t\right) $ instead of $K\left( t\right) $
in Eq. (\ref{a1}). But this is hard to justify. After such a replacement,
for the two junction system, where $K\left( t\right) =1$ (in this case Eq. (%
\ref{a1}) is valid and gives a non-zero current consistent with previous
works), one would get just zero current after the substitution $K\rightarrow %
\mbox{Im}\,K\left( t\right) $. The authors of I, II emphasize explicitly
(between Eqs. (12, 13) in I and Eqs. (8, 9) in II) that Eq. (\ref{a1})
written with $K\left( t\right) $ (not with $\mbox{Im}\,K\left( t\right) $)
allows them to recover the results for the two junction systems.\textbf{\ }

So, we see that following the calculational scheme of I, II one comes
inevitably to the unphysical results, Eqs. (\ref{a7}, \ref{a10}).\textbf{\ }%
In fact, this is a result of one more mistake made by the authors.\textbf{\ }%
It is clear that the non-zero \textquotedblleft
supercurrents\textquotedblright\ in the \textquotedblleft
superinsulator\textquotedblright\ originate from the non-zero value of $%
\mbox{Im}\,K\left( t\right) $\ obtained for the real time $t.$ However, it
follows from the definition of the function $K\left( t\right) $,\ Eq. (\ref%
{a2}), that its imaginary part is identically zero because the Lagrangian $%
L_{0}$ is an even function of $\chi $, whereas the imaginary part of the
exponential entering Eq. (\ref{a2}) is odd in $\chi $. Any type of averaging
with $L_{0}$ gives in this situation zero for $\mbox{Im}\,K\left( t\right) $.

Actually, the non-zero value of $\mbox{Im}\,K\left( t\right) $\ obtained in
I, II is a consequence of an erroneous analytical continuation from the
imaginary time $\tau $ to the real one $t$. The standard procedure of the
analytical continuation for computation of time-dependent quantities is as
follows: i) to calculate the proper correlation function at Matsubara
frequencies, ii) to make analytical continuation to real frequencies, and
then, iii) to perform Fourier transformation to real times. Simple replacing
$\tau \rightarrow it$,\ as was done in I, II, is not correct procedure.
\textbf{\ } \textbf{\ }

Now we review few of \textquotedblleft qualitative
arguments\textquotedblright\ given in I and II.\newline
1) In the last paragraph of left column of p.3 of II, as well as in various
locations in I, the authors write that they obtained \textquotedblleft
global phase-synchronized state...\textquotedblright\ after they used the
model containing no Josephson couplings inside the array. On the other hand,
the authors are aware of the duality relation between the charge and phase
variables (cf. left column of p. 3 in paper I), which means the following
statement: the existence of strong phase correlations implies weakness of
charge number correlations, and vice verse. We are not able to understand
the logics of applying (simultaneously) both the concepts of the charge
quantization (which is the essence of the Coulomb blockade) and the
\textquotedblleft synchronization of phases\textquotedblright .\newline
2) On the first page of I we see an estimate $E_{c}\sim \Delta /g,$ where $%
g\geq 1$ is a dimensionless intergrain conductance, i.e. $E_{c}$
is smaller than the superconducting gap $\Delta $ and thus smaller
than the Josephson coupling $E_{J}\sim g\Delta $. We found no
obvious arguments that would allow one\textbf{\ }to obtain this
estimate in a model which explicitly assumes $E_{c}\gg E_{J}$, as
we see in \textquotedblleft Methods\textquotedblright\ part of I.
\newline 3) The statement in the $1$st paragraph of the right
column, p. 3 of II reads: \textquotedblleft ... even large...
fluctuations in $E_{c}$, $E_{c0}$ and $E_{J}$ as well as offset
charges are negligible as compared to the huge magnitude of
$\Delta _{c}$\textquotedblright . We explain below that if any
threshold voltage $V_{T}=\Delta _{c}/e$ exists, it must scale with
the length of the system, irrespective of any model details. Thus,
the statement cited above contains a comparison of
\textit{intensive} (size-independent) quantities with
\textit{extensive} ones. To give an example: if such a statement
could be correct, any local disorder would be irrelevant to the
thermodynamics properties of any macroscopic system. \newline 4)
The first paragraph of p.4 in paper II contains a comparison
between the lowest-temperature threshold voltage $V_{T}$ and
activation gap $T_{0}$ measured in~\cite{BaturinaExp} at higher
temperatures. In order to \textquotedblleft explain" the high
ratio $eV_{T}/T_{0}$ the authors invoke an idea of
\textquotedblleft dielectric breakdown" and use for 2D system
results of calculations obtained for 1D model. They did it since
for their 2D model they obtained threshold voltage $V_{T}=\Delta
_{c}/e$ that scales just logarithmically with $L$, cf. second line
of (\ref{e2}), which is not enough to get the very large ratio
$eV_{T}/T_{0}$.

We note that: i) nothing in the model studied by the authors indicates the
phenomenon of the\textbf{\ }\textquotedblleft dielectric breakdown" raised
by the authors, ii) there is no reason to invoke such a \textquotedblleft
concept" since the scaling of the threshold voltage with the length $L$ is a
direct consequence of any \textit{correct} calculation leading to a nonzero $%
V_{T}$. The reason for that is simple: the relevant intrinsic parameter for
a macroscopic system is the electric field $\mathcal{E}$ and a critical
value $\mathcal{E}_{T}$ of this field determines the threshold. The value of
$\mathcal{E}_{T}$ does not depend on $L$ for large $L$, whereas the voltage $%
V_{T}\propto L$. The authors of I,II failed to obtain such scaling in 2D due
to several calculational mistakes analyzed above. \newline
5) At last, we mention an explanation given in II, cf. Eq. (14), for the
non-monotonic dependence of the insulating gap on magnetic field: it is
based on a replacement of the disordered superconducting media by a \textit{%
\ single (!) dc SQUID}, with its periodic dependence on magnetic flux
through the SQUID loop. Remarkably, this \textquotedblleft explanation"
disappears completely from paper I, being replaced there by the statement
that the magnetic field suppresses the \textquotedblleft superinsulating
state\textquotedblright\ like it does with the superconducting state. That
version sounds better in view of the experimental data presented in Fig.2,
but it has no relation to the calculations based on the JJ array model. In
addition, it does not give any hint on the non-monotonic $B$-dependence
observed in~\cite{Murthy,BaturinaExp,VBnature} and \textquotedblleft
explained\textquotedblright\ in II.

\textbf{3.} Now we remind the readers some known properties of the model of
I, II derived in a number of earlier papers~\cite%
{Efetov80,Mooij90,MPAFisher90,FazioSchoen91,FeigKorPug97}, cf. also a review
article~\cite{FZ}: \newline
a) The insulating regime is realized at low temperatures if $E_{J}<aE_{c}$,
where $a\sim 1$. In this case the adequate description of charge transport
involves rare hopping of Cooper pairs between the islands, as was explained
already in Ref. \cite{Efetov80}.

The number of these charge excitations is exponentially low at $T\ll E_{c}$,
leading to the exponential suppression of the conductivity. The Coulomb
interaction between the Cooper pairs in the grains was described in Ref.
\cite{Efetov80} using a general capacitance matrix $C_{ij\text{ }}$ without
making any assumption about its parameters. This is why it was assumed that
the energy of adding a particle into the system was of the order $E_{c}$.

b) The model with the vanishing self-capacitance, $C_{0}=0$, employed by the
authors of I, II is special because the energy of interacting charges
\begin{equation}
E_{ch}=\frac{1}{2}\sum_{i,j}\left( C^{-1}\right) _{ij}\hat{\rho}_{i}\hat{\rho%
}_{j}  \label{e13}
\end{equation}%
where $\hat{\rho}_{i}$ are charge operators for the Cooper pairs, is at
large scales linear in 1D and logarithmic in 2D. This is precisely at the
origin of the \textquotedblleft big gap\textquotedblright\ $\Delta _{c}$
found in I, II and displayed in the original form in Eq. (\ref{e2}). Note,
however, a wrong coefficient in the second line for 2D: in fact, the energy
of a single $2e$ excitation is equal to $\frac{2}{\pi }E_{c}\ln (L/d)$. The
logarithmic interaction of charges leads to a charge-binding transition of
the Berezinsky-Kosterlitz-Thouless (BKT) type. The temperature $T_{c}$ of
this transition was estimated~in Refs. \cite{Mooij90,FazioSchoen91} and \cite%
{FeigKorPug97}. Upon approaching this temperature from above, the linear
resistivity diverges, cf. e.g. Eq. (9) of Ref.~\cite{FeigKorPug97}.\newline
c) At $T<T_{c}$ all the charge excitations are paired and linear
conductivity vanishes, in \textquotedblleft dual analogy\textquotedblright\
\cite{MPAFisher90} with the vanishing of the linear resistivity in 2D
superconductors below BKT transition. Nonlinear $I(V)$ transport at low
voltage is due to electric-field induced unbinding of neutral
\textquotedblleft charge molecules\textquotedblright ,
\begin{equation}
I(V)\propto (\mathcal{E}/\mathcal{E}_{T})^{\alpha }\,\quad \quad \alpha
\propto T_{c}/T\,\qquad \mathcal{E}=V/L,  \label{IV}
\end{equation}%
again similar to $V(I)$ characteristics in 2D superconductors below BKT
transition; the threshold electric field can be estimated as $\mathcal{E}%
_{T}\sim E_{c}/ed$. The corresponding characteristic voltage $V_{T}=L%
\mathcal{E}_{T}$ scales with $L$ for trivial reason, being extensive
variable (just like the total current through a superconductive 2D array
scales with its width $L_{\perp }$), without any \textquotedblleft quantum
synchronization\textquotedblright\ \textquotedblleft discovered" in paper I.

\textbf{4.} We have arrived at the conclusion\textbf{\ }that the results
obtained in I, II and expressed by Eqs. (\ref{e1a}, \ref{e1}) are incorrect.
One may wonder if, however, some qualitative features of them might still be
correct. Of a particular interest is the\textbf{\ }\textquotedblleft
superinsulation" formula represented by the low-$V$ limit of Eq.(\ref{e1a}):
\begin{equation}
R\propto \exp \left[ \frac{\Delta _{c}}{E_{c}}\exp (E_{c}/2T)\right]
\label{e15}
\end{equation}%
Since this formula was suggested in paper I for 1D case only, we analyze
this specific case below and show that Eq.(\ref{e15}) is incompatible with
simple physical arguments.

In 1D case $\Delta _{c}\propto L$, and thus, Eq. (\ref{e15}) demonstrates
\textit{both} the exponential dependence on the system size and much faster
than exponential dependence on temperature. If such a dependence on $T$ and $%
L$ could indeed be obtained, it would be really unexpected. We show now that
such a behavior does not exist.

The ground state energy is achieved when there are no charges in the system.
Then Eq. (\ref{e13}) gives zero Coulomb energy $U=0$. A configuration with
the lowest non-zero energy is a dipole consisting of the charges $+2e$ and $%
-2e$ located at neighboring grains. In the model considered here the energy $%
U\left( x\right) $ of the dipole of the length $x$ linearly depends on $x$
and, subjected to the external electric field $\mathcal{E}$, equals
\begin{equation}
U\left( x\right) =2e\left( \mathcal{E}_{c}-\left\vert \mathcal{E}\right\vert
\right) x  \label{e16}
\end{equation}%
The energy of the dipole linearly grows with $x$ for $\left\vert \mathcal{E}%
\right\vert <\mathcal{E}_{c}$. A current through the system is possible
provided the size of the dipole can reach the size of the system $L.$ The
energy corresponding to such a dipole is huge:
\begin{equation}
U\left( L\right) =2\left( \Delta _{c}-eV\right) =2eL(\mathcal{E}%
_{c}-\left\vert \mathcal{E}\right\vert )  \label{e17}
\end{equation}%
where $\mathcal{E}_{c}=\Delta _{c}/eL$. However, the probability of creating
this dipole is finite at finite $L$ and proportional to $\exp (-U(L)/T)$,
therefore the resistivity $R$ can be written as
\begin{equation}
R\propto \exp \left[ \frac{2eL\left( \mathcal{E}_{c}-\left\vert \mathcal{E}%
\right\vert \right) }{T}\right]  \label{e18}
\end{equation}%
We see from Eq. (\ref{e18}) that the resistivity obtained from this simple
consideration contains the combination $\Delta _{c}-eV$ and not square of
it. Therefore, the Coulomb blockade is important for $eV<\Delta _{c}$ only.

Eq. (\ref{e18}) shows that the Coulomb blockade can be overcome if one
creates a dipole with the energy of order $\Delta _{c}$. Clearly, this
probability is small but it is described by the activation law. This simple
argument excludes the double-exponential temperature dependence of the
resistivity $R$, Eq. (\ref{e15}).

We want to emphasize that the big gap in the spectrum of the excitations $%
\Delta _{c}$ exists (within the specific 1D model discussed) not only for
the Cooper pairs but for the normal electrons as well \cite{ET} (in the
latter case one would have to use $\Delta _{c}/4$ instead of $\Delta _{c})$.
So, if the \textquotedblleft superinsulation\textquotedblright\ was due to
an arbitrarily chosen special form of the capacitance matrix, it would exist
even without any superconducting pairing.

An important remark is in order: following I and II we considered the
situation of an \textit{exponential} dependence of resistivity on the system
size $L$. In reality, this usually does not make sense for large systems: in
2D case a resistance \textit{per square} $\rho (T)$ can be defined, whereas
in 1D case the resistance grows $\propto L$. This behavior is due to
inelastic processes including thermal activation. Such processes (neglected
within the calculational scheme used in papers I and II) lead, in
particular, to the current-voltage dependence (\ref{IV}) for 2D case. 1D
model with weak $E_{J}$ and vanishing self-capacitance is very specific due
to its simplicity: it reduces to a single capacitor with $C_{\mathrm{eff}%
}=C/N$. Thus, the meaning of Eq. (\ref{e18}) is also trivial: it corresponds
to a resistance through a capacitor in the Coulomb blockade regime.
Actually, non-trivial macroscopic systems with a low-temperature
conductivity proportional to $\exp \left( -cL\right) $ do exist: such a
behavior is a consequence of a \textquotedblleft quantum topological order",
cf. e.g.~\cite{IoffeFeigelman2002}. The standard 2D Josephson-junction model
discussed in I, II does not lead to such a behavior.

\textbf{5.} Could this 2D regular JJ array model 
be used to describe experimental data~\cite{Murthy,BaturinaExp} on
homogeneously disordered superconductive films? The authors of I wrote (top
of the last column of the main text): \textquotedblleft our understanding of
the origin of superinsulating state in the films relies on the formation of
the network of superconducting droplets within the normal
matrix\textquotedblright . Still, the model they considered contains
superconductive islands connected by tunnel barriers without any sign of
normal matrix inside. The problem of the superconductive islands put into a
normal matrix is essentially different in its physics and more complicated
for theoretical studies~\cite{FL,Spivak,FLS,SF} in comparison to the model
of 2D JJ array; its relevance to the data of~\cite{Murthy,BaturinaExp} could
be a subject of separate work. Coming back to the discussed model of 2D JJ
array, we note its serious deficiencies regarding a possibility of an
application to homogeneously disordered superconductors: \newline
i) in the absence of well-defined (structural) grains separated by tunnel
junctions, the notion of the charging energy $E_{c}$ becomes ill-defined, to
begin with;\newline
ii) even if the grains separated by tunnel barriers are assumed to appear
\textquotedblleft somehow\textquotedblright\ in a continuously disordered
media, there is no reason to expect them to form a regular lattice with
identical areas, $E_{J}$ and $E_{c}$. \newline
iii) even if in spite i) and ii) a granular model could with some degree of
imagination be thought of as relevant, the model considered in I, II
contains a vanishing ratio of the capacitances $C_{0}/C$ and this is the
crucial point to obtain the \textquotedblleft huge\textquotedblright\ gap;
why should this condition be fulfilled in a \textquotedblleft
self-organized\textquotedblright\ granular network? \newline
iv) care should be exercised with respect to the role of random stray
charges, which are always present in real arrays and influence strongly the
insulating part of the phase diagram for any model where the insulating
behavior is due to Coulomb repulsion (cf. e.g.~\cite{GK,PF} and references
therein). In particular, in the presence of a charge disorder with a typical
amplitude of order $e$, the charge-pairing BKT-like transition does not
exist.

Based on these arguments we question any kind of \textquotedblleft
fitting\textquotedblright\ the experimental data~ \cite{BaturinaExp} by
results of \textit{any} calculations based on the simplified JJ array model.

To summarize: 
the phenomenon dubbed as \textquotedblleft quantum
synchronization\textquotedblright\ does not exist in the model studied. This
model describes the well known phenomena of Coulomb blockade. The result
obtained by the authors are in conflict with previous studies of the model
and we believe are based on erroneous calculations.



\bigskip

\section{Note added on July 30, 2008.}

Recently, M. V. Fistul, V. M. Vinokur, T. I. Baturina (FVB) submitted a more
detailed paper III (\textit{Macroscopic Coulomb blockade in large Josephson
junction arrays, }arXiv:0806.4311, v1). Several formulae criticized in our
comment have been corrected. In III the current $I_{s}\left( V\right) $ is
expressed through a retarded correlation function of hopping amplitudes and
it is shown how to write this function at real time. This makes the current
an odd function of the voltage $V$ as it should be.

However, the most serious errors remain. A major approximation
when the first cosine in Eq. (9) (of our Comment) is replaced by a
constant would be justified if the Josephson couplings in the bulk
were large. However, the authors consider an opposite limit. There
the phase difference between grains in the bulk strongly
fluctuates with time and the stationary perturbation theory used
in III, Eqs. (10-13), is not applicable. For the same reason, the
arguments lead to Eq. (31) of III are not consistent. Setting all
couplings in the bulk to zero, as the authors do, breaks contacts
between grains. This system (thus approximated by a capacitor)
cannot support a finite \textit{dc }current.

Nevertheless, in papers I-III the authors present formulae for
\textit{dc }current as a smooth function of voltage $V$ (see Eqs.
(46, 54) of III). How could it happen?

We have found that an integral obtained as a result of their approximation
has not been calculated properly.

Following I-III one can write the current in the form of Eq. (25)
of III with the correlation function $K\left( t\right) $ written
in the first equation of Eqs. (33) of III (or, Eq. (52) of III).
For a large number of grains one can neglect an influence of the
leads and put $K_{leads}\left( t\right) =1$. The integral obtained
in this way can be easily evaluated in different limits.

For example, in the limit $E_{c}/\ln N\ll T\ll E_{c},$ where $N$ is the
number of the grains in the system, the authors of III obtained their Eq.
(53). They calculate the integral using the saddle point method: expanding $%
\cos \left( E_{c}t\right) $ entering Eq. (53) in $t$ near $t=0$
and then computing the Gaussian integral around this saddle point.
In this way the authors obtained the double exponential behavior
(see Eq. (1) of our comment).

However, $t=0$ is not the \textit{only} saddle point of this integral: all $%
t=\pm 2\pi m$ with integer $m$ are saddle points as well, and all of them
must be taken into account. Integrating near all the saddle points and
summing them up yields the following expression for the current
\begin{equation}
I_{s}\left( V\right) =\frac{\pi ^{3/2}B}{\sqrt{N}}e^{E_{c}/4T}\left(
I_{0}\left( V\right) -I_{0}\left( -V\right) \right) ,  \label{k1}
\end{equation}%
where
\[
I_{0}\left( V\right) =\sum_{k=-\infty }^{\infty }\delta \left(
2eV-E_{c}\left( N/2+k\right) \right) \exp \left( -\frac{k^{2}}{4N}%
e^{E_{c}/2T}\right)
\]%
and $B$ is a coefficient proportional to $E_{J}^{4}$.

Eq. (\ref{k1}) contains a sum of matrix elements between the
states with different numbers of the particles of \emph{isolated
}grains. The discreteness of the levels is the consequence of the
discreteness of the charge of the Cooper pairs. It is there at all
temperature regimes once contacts are broken. The levels may be
smeared only by taking into account charge tunnelling between the
grains which has been explicitly ignored by the approximation made
by the authors. A tunnelling (a non-zero $E_{J}$) smoothes $I(V)$
functions, not the temperature. This, seems to be an obvious
point, has been discussed in the main part of our Comment (see
e.g., Eqs. (18,22)). No disorder can help in this respect as long
as the contacts remain broken.

Correct calculation of the integral does not save Eq. (\ref{k1}),
though. It still shows a dc-current flowing through a capacitor at
resonant values of voltage.

Finally we comment on the last paragraph on p.11 of III. There the authors
explain how one should think of \textquotedblleft superinsulating" state:
the internal part of the array (except the rightmost and leftmost junctions)
\textquotedblleft acts coherently as a single superconducting island". It is
not quite clear whether the authors have in mind a closed ring to be a
superconductor or \textquotedblleft superinsulator". Regardless of this, we
do not think that superconductivity in an array with broken junctions is
possible anyway.

We conclude that the new theoretical material presented in III neither
clarifies nor essentially corrects the analysis and the results of I and II.

\end{document}